\documentclass[twoside,journal]{IEEEtran}
\usepackage{cite}
\usepackage{amsmath,amssymb,amsfonts}
\usepackage{algorithmic}
\usepackage{graphicx}
\usepackage{textcomp}
\usepackage{wrapfig}
\usepackage{filecontents}
\usepackage{graphicx,import}
\graphicspath{{figures/}}
\usepackage{xcolor}

\newcommand{\R}{\mathbb{R}}

\renewcommand{\Re}{\text{Re\,}}
\renewcommand{\Im}{\text{Im\,}}

\newcommand{\im}{\textnormal{im}}
\newcommand{\re}{\textnormal{re}}

\newcommand{\jw}{j\w}
\newcommand{\wk}{\w_k}

\newcommand{\w}{\omega}
\newcommand{\W}{\Omega}

\newcommand{\norm}[1]{\Vert {#1} \Vert}
\newcommand{\abs}[1]{\left| #1\right|}

\newcommand{\Csp}{\mathcal{C}}

\newcommand{\Dsp}{\mathcal{D}}

\newcommand{\Fsp}{\mathcal{F}}

\newcommand{\Ksp}{\mathcal{K}}

\newcommand{\Nsp}{\mathcal{N}}
\newcommand{\Psp}{\mathcal{P}}
\newcommand{\Rsp}{\mathcal{R}}

\newcommand{\bx}{\boldsymbol{x}}

\newcommand{\bZ}{\boldsymbol{Z}}

\newcommand{\btheta}{\boldsymbol{\theta}}

\newcommand{\bmu}{\boldsymbol{\mu}}

\newcommand{\tbmu}{\tilde{\bmu}}

\newcommand{\ZDRT}{Z_{\text{DRT}}}
\DeclareMathOperator*{\argminC}{\arg\min}

\begin{document}
%
\title{Time-Constant-Domain Spectroscopy: An Impedance-based Method for Sensing Biological Cells in Suspension}

\author{Roberto~G.~Ram\'irez-Chavarr\'ia, Celia~S\'anchez-P\'erez, Luisa Romero-Ornelas, Eva Ram\'on-Gallegos
\thanks{R.G. Ram\'irez-Chavarr\'ia is with Instituto de Ingenier\'ia, Universidad Nacional Aut\'onoma de M\'exico, 04510, Ciudad de M\'exico, M\'exico (e-mail: RRamirezC@iingen.unam.mx). }
\thanks{C. S\'anchez-P\'erez is with  Instituto de Ciencias Aplicadas y Tecnolog\'ia, Universidad Nacional Aut\'onoma de M\'exico, AP 10-186, 04510, Ciudad de M\'exico, M\'exico (e-mail: celia.sanchez@icat.unam.mx).}
\thanks{L. Romero-Ornelas and E. Ram\'on-Gallegos are with 
	Escuela Nacional de Ciencias Biol\'ogicas, Campus Zacatenco, Instituto Polit\'ecnico Nacional, 07738, Ciudad de México, M\'exico (e-mails: crakfdash@hotmail.com; evaramong72@gmail.com).}
\thanks{Manuscript received July, 2020; revised xxx xx, 20xx.}}

\markboth{IEEE Sensors Journal,~Vol.~xx, No.~xx, Month~20xx}{Ram\'irez-Chavarr\'ia \MakeLowercase{\textit{et al.}}: Time-Constant-Domain Spectroscopy: An Impedance-based Method for Sensing Biological Cells in Suspension }
%



\maketitle

\begin{abstract}
Impedance measurement is a common technique to characterize and detect the electrical properties of biological cells. However, to decode the underlying physical processes, it requires complex electrical models alongside prior knowledge of the sample under study. In this work, we introduce an attractive label-free method for sensing biological cells in suspension based on the measurement of electrical impedance and the distribution of relaxation times (DRT) model. The DRT maps impedance data from the frequency-domain to a time-constant-domain spectrum (TCDS) being a useful and robust method for data analysis. We perform impedance measurements in the range from 1 kHz to 1 MHz to obtain the TCDS for sensing mimic samples as well as HeLa cells in suspension. Results show that the TCDS can be seen as an electrical fingerprint for the sample, as it can decode useful information about the composition and structure with high sensitivity and resolution.
\end{abstract}


\IEEEpeerreviewmaketitle

\section{Introduction}
{\label{sec:Intro}}
\IEEEPARstart{T}{he} measurement of electrical/electrochemical impedance is a powerful tool for investigating electric properties of biological cells suspended in a conductive medium~\cite{Barsoukov2005,Asami2002}. When these suspensions are exposed to an external electric field $\vec{E}(\w)$, at an arbitrary frequency $\w$, the cells are polarized, accumulating electrical charge at the interface of its shell and the surrounding medium. The polarization $\vec{P}(\w)$ depends on the nature of the sample as well as on the frequency~\cite{Feldman2005}. In general, the polarization does not reach instantaneously its steady-state, and it takes some time to reach its final value. Furthermore, when $\vec{E}$ is suddenly removed, the polarization exponentially decays after a certain elapses~\cite{raicu2015dielectric}. This process is the so-called dielectric relaxation phenomenon, given by a relaxation time-constant $\tau$. The electric response of cell suspensions is mainly supported on the effective medium theory and the Maxwell\textquotesingle s mixture model~\cite{Grimnes2002}. These models relate the intrinsic electrical properties of the suspension, the cell, the suspending medium and the volume fraction. Coupled with the physics-based models, in practice, one needs analytical techniques to assess the electrical response of the sample. Particularly, for biological cells, the electrical impedance spectroscopy (EIS)~\cite{Lasia2014} is widely accepted as a label-free, non-invasive and quantitative analytical method that can thoroughly assess the electrical properties of biological cells~\cite{Zhu20101}. 


Measuring electrical impedance of biological samples is mainly intended to estimate the characteristics of an analyte such as cells~\cite{Mansoorifar2018}, bacteria~\cite{Clausen2018} and proteins~\cite{HALLIWELL2014}.
Particularly, for a cell culture, the aim of impedance characterization is to provide a method to obtain biological parameters and assess different scenarios some of which are, motility, cell growth and identification, and cell counting, to mention only a few. In this sense, a challenging task is to experimentally increase accuracy and to enhance the detection sensitivity. Some approaches have been proposed to compensate, for instance, undesirable effects which commonly shunts the electrical processes\cite{Ishai_2013}. Moreover, the detection sensitivity can be outperformed by a specific geometry of the electrodes~\cite{Chang2016} and using microfluidic technology~\cite{ILIESCU2007}. Also, by either improving the electrode material~\cite{Koutsouras2019} or by modifying the electrode surface~\cite{Hwang2019}, it increases the detection limits and the resolution. More recently, advances in sophisticated instrumentation~\cite{Xiaowen2014,Carminati2017} and detection methods~\cite{Spencer2020}, have resulted in promissory tools for impedance measurement. Once experimental data are collected, complex electrical models are required to decode the underlying electrical processes. Such a task is, in general, performed by fitting the impedance data to the well-known Cole-Cole model~\cite{Iglesias2017} and lumped electrical equivalent circuits~\cite{VYROUBAL2018}, or in some cases, by comparing experimental data with rigorous theoretical models~\cite{Sun2007}. This task, however, is not always trivial, and requires some prior knowledge of the sample under study to identify the underlying processes. Even though this semi-empirical methods are widely accepted, they are mainly limited to quantitative analysis, and are not able to capture the detail of the complete electrical properties at high throughput~\cite{MEDEIROS2016,Spencer2020}. 

On the other hand, the distribution of relaxation times (DRT) model~\cite{Boukamp2015} presents an alternative approach for interpreting electrical impedance data through the relaxation theory. It can isolate and identify all the relaxation processes from an EIS spectrum, by transforming data from the frequency-domain $\w$ into a time-constant-domain $\tau$ given by a distribution function. Thus, one can systematically build a suitable and intuitive impedance model for qualitative data evaluation and interpretation. Despite its narrow relationship with the electrical properties of cell suspensions~\cite{Sun2010}, to the best of our knowledge, the DRT model has been barely studied for analyzing EIS data from biological cells. 

In a previous work~\cite{RamirezSAS}, we showed preliminary ideas on electrical impedance measurements and DRT analysis for detecting the concentration of microparticles in a colloidal suspension. In this paper, we describe an attractive method that characterizes biological cell suspensions with the following main contributions. We introduce first results of the so-called time-constant-domain spectroscopy (TCDS), as an attractive way to thoroughly characterize biological cells suspended in a highly conductive medium, used for culture. The TCDS can be seen as an electrical fingerprint for the cell suspension, representing a systematic procedure for estimating its concentration. Ultimately, from TCDS, a simple parametric impedance model can be derived to decode the electrical properties of biological cells in suspension.

The rest of the paper is organized as follows:
Section~\ref{sec:Theory} introduces the useful concepts to describe the TCDS method by electrical impedance measurements and the DRT model. The details of the measurement setup are presented in Section~\ref{sec:System}. The experimental results are shown along with a thorough discussion in Section~\ref{sec:results}. Finally, Section~\ref{sec:conclusions} is devoted to the conclusions.

\section{\label{sec:Theory}Method description}
\subsection{\label{sec:EIS}Electrical impedance}
Let us define the discrete Fourier transform $\Fsp$ of the applied voltage $V(\w_i)\triangleq\Fsp\{u(t)\}$ and the measured current $I(\w_i)\triangleq\Fsp\{y(t)\}$. The electrical impedance is given by the ratio of voltage and current phasors at an arbitrary frequency $\w_i$, for $i=1,2,\dots,N$. The $i-$th impedance point is given by
\begin{equation}
\label{Z}
Z(j\w_i)\triangleq \frac{V(\w_i)}{I(\w_i)}=\abs{Z}\left(\cos(\phi) + j \sin(\phi)\right),
\end{equation}
where $j=\sqrt{-1}$, $\abs{Z}$ is the impedance magnitude, and $\phi$ is the phase angle between the voltage and current phasors. Also, impedance can be represented by its real $Z^\re(\w)=\abs{Z}\cos\phi$, and imaginary $Z^\im(\w)=\abs{Z}\sin\phi$ components. Throughout this paper, we denote $\hat{Z}(j\w)$ as the measured impedance spectrum, with its associated components $\hat{Z}^\re$ and $\hat{Z}^\im$.

\subsection{\label{sec:equival_circuit}Equivalent circuit models}
According to the literature, when measuring EIS in cell suspensions, the electrical equivalent circuit (EC) is the classical approach for interpreting the resultant spectra as the EC parameters can be related to the composition and electrical processes involved. The impedance model of biological cellular suspensions is given by an EC working in the range from kHz up to tens of MHz~\cite{Qiao2012,Addabbo2019,Ramirez2019Meas} as follows 
\begin{align}
Z_{\text{cir}}(j\omega)= \frac{j\w R_m  C_\text{e}+1}{j\w C_\text{e}} + \frac{R_{\text{p}}}{1+j\w R_\text{p}C_\text{p}},
\label{eq:zcir}
\end{align}
where $R_{\text{m}}$ is the  resistance of the medium, $C_\text{e}$ represents the capacitance of the electrodes,  $R_\text{p}$ corresponds to the resistance of the dielectric particles, and $C_\text{p}$ is their capacitance. The relaxation times in this circuit are given by $\tau_1=R_{\text{m}}C_\text{e}$ and $\tau_2=R_\text{p}C_\text{p}$.
In practice, the capacitors $C_\text{e}$ and $C_\text{p}$ may have a non-ideal behavior and are reformulated as constant-phase-elements (CPEs) with impedance given by
\begin{align}\label{eq:Zcpe}
Z_{\text{CPE}}=\frac{1}{\left(j\omega\right)^\alpha Q},
\end{align}
where $Q$ is the non-ideal capacitance and $\alpha \,\in\left(0,1\right]$ is an exponent tunning the non-ideality. By combining~\eqref{eq:Zcpe} and~\eqref{eq:zcir}, one obtains the impedance expression for the circuit with CPEs. Note that, if $\alpha=1$ in~\eqref{eq:Zcpe}, it resembles the impedance of an ideal capacitor. Though the CPE provides a convenient model, even when the nature of the system is unknown, it does not have any particular interpretation for identifying the underlying electrical processes~\cite{alim2017}.

\subsection{\label{sec:DRT}Distribution of relaxation times model}
The distribution of relaxation times (DRT) is a useful approach to analyze EIS measurements, commonly used in electrochemical applications, but almost never used in biological cell suspensions. By using DRT, it is possible to obtain a distribution function that joins up the polarization and conduction processes involved in the sample under study. The impedance associated to the DRT model is given by a Fredholm integral of the first-kind as
\begin{equation} \label{eq:ZDRT}
Z_{\text{DRT}}(j\omega)=R_{\infty}+\int_0^\infty \frac{\gamma(\ln\tau)}{1+j\omega\tau}d(\ln\tau),
\end{equation}
where $R_{\infty}$ is the high-frequency resistance, $\gamma(\ln\tau)$ is the distribution function, and $\tau$ is the relaxation time scale. Solving~(\ref{eq:ZDRT}) for $\gamma(\cdot)$ is, however, an inverse problem with an infinite number of solutions~\cite{Groetsch_2007}. To solve the integral equation in~\eqref{eq:ZDRT} several techniques have been applied, such as regularized least-squares~\cite{RAMIREZSYSID, Saccoccio2014}, algebraic reconstruction~\cite{BuschelDRT} and Bayesian inference~\cite{Ciucci2015}. Therein, the aim is to find the optimal $\gamma(\ln \tau)$ function, such that $\hat{Z}(j\w)\approx Z_{\text{DRT}}(j\w)$. Once the $\gamma(\cdot)$ function is retrieved, it can be seen as a time-constant-domain spectrum (TCDS) given by a plot of $\gamma(\cdot)$ against $\tau$, and hence, different relaxation mechanisms can be easily isolated and identified in this spectrum.

\subsection{Time-constant-domain spectrum estimation}
The problem of solving the DRT model leads to the estimation of the TCDS. To this end, we implemented ridge regression for solving a quadratic programming optimization problem as follows. The distribution function $\gamma(\ln\tau)$ in~\eqref{eq:ZDRT}, can be approximated by the sum of $M$ radial basis functions (RBFs) $g_m(\ln\tau) \triangleq \exp{-(\mu (|\ln\tau-\ln\tau_m|)) ^2}$, where each function is centered at the $m-$th relaxation time $\tau_m$, with a full width at half maximum (FWHM) given by $\mu$.  
Hence, a parametrization of $\gamma(\cdot)$ is given by  $\hat{\gamma}(\ln\tau; \btheta) \triangleq \sum_{m=1}^{M} \theta_m g_m(\ln\tau)$, where $\btheta=[\theta_1 \quad \cdots \quad \theta_M]^{\top}$ is a vector of parameters that relates the amplitudes of the RBFs. The DRT model is thus written in terms of the parameter vector $\btheta$ and the basis functions as follows 
\begin{equation}\label{eq:ZDRT_dis}
\ZDRT(j\omega; \btheta)=R_{\infty}+\sum_{m=1}^{M} \theta_m \int_0^\infty \frac{g_m(\ln\tau)}{1+j\omega\tau}d(\ln\tau).
\end{equation}

Multiplying \eqref{eq:ZDRT_dis} by its complex conjugate leads to the real component 
\begin{align}
\ZDRT^\re  =  R_{\infty} + \sum_{m=1}^{M} \theta_m \underbrace{\left[\int_0^\infty \frac{g_m(\ln\tau)}{1+(\w \tau)^2}d(\ln\tau)\right]}_{A^{\textnormal{\re}}},
\label{eq:zdrtre}
\end{align}

and the imaginary component 

\begin{align}
\ZDRT^\im  = - \sum_{m=1}^{M} \theta_m \underbrace{\left[\int_0^\infty \frac{\w \tau g_m(\ln\tau)}{1+(\w \tau)^2}d(\ln\tau)\right]}_{A^{\textnormal{\im}}}.
\label{eq:zdrtim}
\end{align}

The measured real and imaginary impedance components, at $\ell=1, \dots ,L$ frequency points, are given by  $\hat{\bZ}^{\textnormal{\re}}\in \R^{L\times 1}$ and $\hat{\bZ}^{\textnormal{\im}}\in \R^{L\times 1}$, respectively.  Furthermore,~\eqref{eq:zdrtre} and~\eqref{eq:zdrtim} can be written in matrix notation as $\mathbf{A}^{\textnormal{\Re}} \in \R^{L\times M}$ and $\mathbf{A}^{\textnormal{\Im}} \in \R^{L\times M}$, respectively. The optimal parameter vector $\btheta^\star$, in the least squares sense, is retrieved by minimizing the error between the model and measurements plus a penalty term $\Dsp$, as follows
\begin{align}
\begin{aligned}
V(\btheta)=& \norm{\left((R_{\infty} \mathbf{1})+\mathbf{A}^{\textnormal{\re}}\btheta\right) - \hat{\bZ}^{\textnormal{\re}}}_2^2 \; + \; \norm{\mathbf{A}^{\textnormal{\im}}\btheta - \hat{\bZ}^{\textnormal{\im}}}_2^2 \; + \; \Dsp,
\end{aligned}\label{eq:cost}
\end{align}
being $\mathbf{1}\in \R^{L\times 1}$ a vector of ones. The penalty term is given by the $\ell_2$ norm of $\btheta$ as $\Dsp\triangleq \lambda \norm{\btheta}_2^2$, where $\lambda > 0$ is a hyper-parameter tuned on the data~\cite{Chen2013}. The expression in~\eqref{eq:cost} corresponds to the well-known ridge regression form of regularized least-squares. It deals with poorly conditioned equations, prevents over-fitting and imposes smoothness on the solution. Ridge regression is motivated by the following constrained minimization problem, 
\begin{align}
\btheta^\star = \argminC_{\theta:\theta\geq 0} V(\btheta), \label{eq:min}
\end{align}
which is said to be a convex optimization problem with a closed form solution, for which $\btheta^\star$ is the optimal parameter vector.

Once $\btheta^\star$ is computed, the TCDS is directly given by the distribution function  $\hat{\gamma}(\ln\tau;\btheta^{\star})$. According to the relaxation theory, the expected process analytically exhibits a Gaussian-like distribution function with local maximum centered at its corresponding relaxation time $\tau_m$. It is hence possible to state that, the number of maxima in the TCDS is related to the number of physical processes in the measured impedance spectrum. This is the major potentiality of our method as it is able to find the characteristic relaxation times of suspensions in a systematic way. Moreover, the TCDS can be seen as a series connection of parallel-$RC$ networks, each one related with an electrical process characterized by the $i-$th relaxation time-constant $\tau_i=R_iC_i$.  As a result, it is possible to relate the TCDS with an equivalent circuit where the number of maxima in $\gamma(\cdot)$ is equal to $n$ parallel $RC$ networks connected in series. Ultimately, the identified TCDS-based equivalent circuit has a model structure given by
\begin{align}\label{eq:model_drt}
H_{\text{TCDS}}(j\omega)=R_{\infty} + \sum_{i=1}^{n} \frac{\Rsp_i}{1+j\omega\tau_i},
\end{align}
where $\Rsp_i$ is a positive definite constant proportional to the amplitude of the distribution function.

\section{Measurement setup}
\label{sec:System}
\begin{figure}[t]
	\centering
	\includegraphics{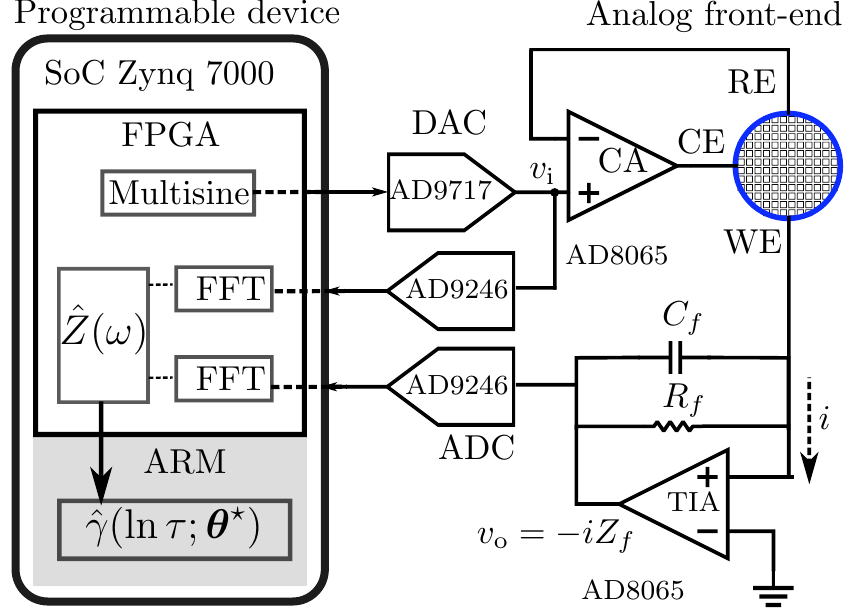}
	\caption{Block diagram of the setup for measuring the impedance $\hat{Z}(\omega)$ and computing the TCDS $\hat{\gamma}(\ln\tau;\btheta^{\star})$}.
	\label{fig:system}
\end{figure}

The measurement setup is shown in Fig.~\ref{fig:system}. It comprises a programmable device with dedicated features  for EIS measurement and DRT processing, an analog front-end for data conditioning alongside a potentiostat circuit, and an electrochemical cell as the sensing element. The FPGA stores data of the excitation signal, which is converted into its analog form by a 14-bit digital to analog converter (DAC) at a sample rate $f_\text{sDAC}=125$~MHz. The input signal $v_{\text{i}}(t)=\sum_{k=1}^{F} A_k\cos(\omega_kt+\delta_k)$, is a multisine with amplitude $A_k=\pm 5~\text{mV}$, phase $\delta_k$ and $\omega_k=2\pi f_k$ excited frequency. The excitation comprises $F=52$ frequencies within the range from 1 kHz to 1 MHz, i.e. approximately 17 frequencies per decade in a quasi-logarithmic scale. The potentiostat circuit uses the control amplifier (CA) for applying $v_i$ between the counter electrode (CE) and the reference electrode (RE). The CA compensates its output by measuring the voltage drop at the RE, and the transimpedance amplifier (TIA) measures the current $i$ flowing through the sample at the working electrode (WE). The output voltage, $v_{\text{o}}$, is proportional to the current through the feedback parallel impedance $Z_f=R_f//C_f$. Ultimately, a reference of the excitation signal $v_{\text{i}}$ and the output voltage $v_{\text{o}}$, are synchronously digitized by a pair of 14-bit analog to digital converters (ADC) operating at a sampling frequency $f_\text{sADC}=8.19$ MHz.
For low uncertainty impedance measurements, an integer number $P$ of periods of the input $v_{i}(t)$ and output $v_{o}(t)\propto i(t)$ signals are transformed to the frequency-domain by the fast Fourier transform (FFT). The input/output spectra are defined as $V_{\text{i}}(k)\triangleq \Fsp \{v_{\text{i}}(t)\}$ and $V_{\text{o}}(k)\triangleq \Fsp \{v_{\text{o}}(t)\}$, respectively. The measured impedance $\hat{Z}(j\w)$, at each single experiment $r$ over $p=1,\dots,P$ periods, is given as the ratio of the FFT sample means
\begin{align}
\hat{Z}^{[r]}(j\w_k)\triangleq
\frac{\frac{1}{P}\sum_{p=1}^{P} V_{\text{i}}^{[p]}(k)}{\frac{1}{M}\sum_{p=1}^{P} V_{\text{o}}^{[p]}(k)/Z_f}.\label{eq:estim}
\end{align}
Averaging over $r=1,\dots,R$ repetitions, a consistent and unbiased non-parametric estimator leads to the measured impedance $\hat{Z}(j\w) =\hat{Z}^\re + j \hat{Z}^\im$. Due to the properties of the spectral averaging technique~\cite{Pintelon2001}, this method increases the signal-to-noise ratio (SNR) by considering a large number of $R$ experiment repetitions in a measurement, thus benefiting the computation of the time-constant-domain spectrum.
Once the impedance data are retrieved, the FPGA streams them to a dual-core ARM\textsuperscript{\textregistered} Cortex\textsuperscript{\textregistered}-A9 processor running a 32-bit Debian GNU/Linux operating system at 800 MHz. A custom application\footnote{The code is available on \texttt{https://github.com/rgunam/TCDS}}, written in Python 3, computes the TCDS given by the $\hat{\gamma}(\ln\tau;\btheta^{\star})$ block in Fig.~\ref{fig:system}. The TCDS application executes elementary matrix operations, and runs an optimization solver~\cite{agrawal2018} for minimizing the cost function in~\eqref{eq:cost}. Regarding the electrochemical cell, we used a
microelectrode array (Dropsens\textsuperscript{\textregistered} DRP-G-MEA555), where CE, RE, and WE are made of platinum (Pt) over a glass substrate. The sensing element,
namely the WE, has a honeycomb structure with a 3 mm diameter surface and 620 micro-electrodes of 10 $\mu$m, separated each by a distance of 10 $\mu$m. It is worth mentioning that the WE structure benefits the measurement by rapidly increasing the amount of steady-state current density, which enhances the sensitivity and detection limits~\cite{Tedjo2019}.   

\section{Results and discussion}\label{sec:results}

\subsection{Validation of the method}
\begin{figure}[t]
	\centering
	\includegraphics[width=1.0\columnwidth,scale=1]{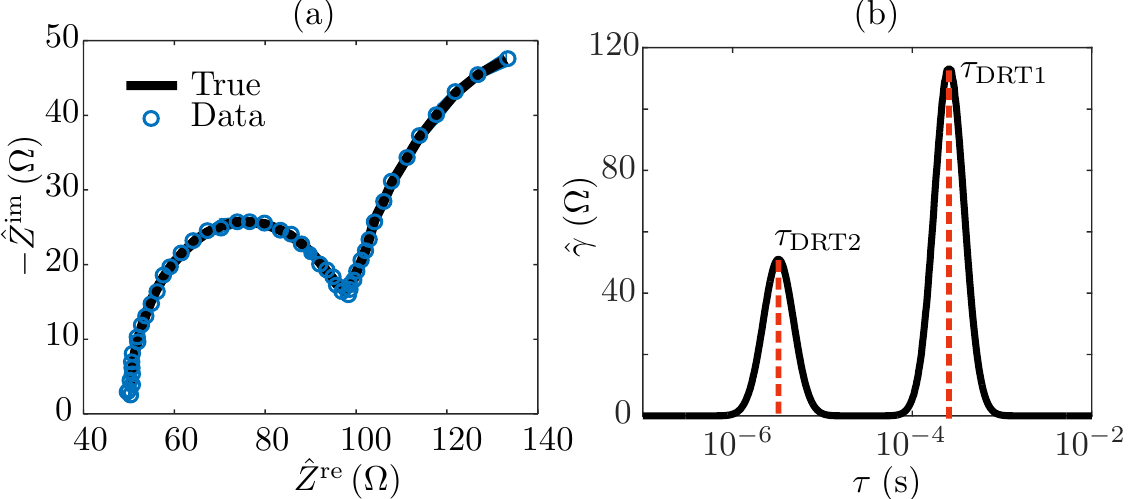}
	\caption{Results of the system validation using an electrical circuit. (a) Nyquist diagram showing the imaginary and real parts relationship. (b) TCDS by means of the estimated $\hat{\gamma}(\cdot)$ function.}
	\label{fig:calibra}
\end{figure}

We experimentally validated the performance of the proposal by an electrical circuit with equivalent impedance as in~\eqref{eq:zcir}. The network was constructed using discrete components with values of $R_{\text{m}}=97.2~\Omega$, $C_\text{e}=8.27~\mu\text{F}$, $R_\text{p}=48.5~\Omega$ and $C_\text{p}=66.4~\text{nF}$, measured with a UNI-T\textsuperscript{TM} UT612 commercial LCR meter (error $<$ 1\%). We performed $R=10$ repetitions of impedance spectra measurements and calculated $\hat{Z}(j\w_k)$ using the procedure described in the Section above. Results are shown in Fig.~\ref{fig:calibra}(a), where we depict the Nyquist diagram for EIS measurements (dots) along with the true impedance spectrum (solid line), calculated using the relation in~\eqref{eq:zcir}. One can see a good agreement between two plots with the maximum standard deviation of 0.54~$\Omega$ for $\hat{Z}^\re$ and 0.69~$\Omega$ for $\hat{Z}^\im$. The Nyquist diagram exhibits two main processes involved, a semicircle on the left side (high frequencies) and a curve on the right side (low frequencies). Hence, it is possible to deduce two relaxation process in the sample, related to the time-constants $\tau_1$ an $\tau_2$. Nevertheless, in the frequency- domain, it is not possible to precisely determine where the $\tau$'s are located.

For the identification of the relaxation times, we used the DRT model to compute the TCDS given by the distribution function $\hat{\gamma}$ shown in Fig.~\ref{fig:calibra}(b). One can clearly distinguish two Gaussian-like peaks with their maxima centered at the characteristic time-constants $\tau_{\text{DRT1}}=2.17$~ms and $\tau_{\text{DRT2}}=3.25~\mu\text{s}$. The actual time-constants of the calibration circuit, calculated by using the nominal values, are $\tau_1=2.13$~ms and $\tau_2=3.22~\mu\text{s}$. By comparing both results, the validation leads to an estimation error as low as 2\%, showing good consistency with the impedance measurements $\hat{Z}$. As expected, the TCDS gives an electrical fingerprint representation of the sample under study by means of the time-constants, from which the complete electrical properties can be extracted at high throughput.

\subsection{TCDS analysis for mimic cell suspensions} \label{Syscal}
As a proof of concept, we used micro-colloidal samples for two main reasons, it is well known that they can mimic cellular suspensions~\cite{XU2016}, and it is possible to get a clear insight of the expected electrical response~\cite{raicu2015dielectric} to verify the performance of the TCDS. The samples were composed by $\text{SiO}_2$ micro-particles with a diameter $d=0.8~\mu\text{m}$, suspended in a conductive phosphate buffer saline (PBS 1X) matrix at seven different concentrations $\kappa$ (wt.\%). According to the literature~\cite{XU2016}, $\text{SiO}_2$ particles exhibits a large dielectric behavior, with permittivity  $\epsilon_{\text{SiO}_2}= 4.5$ and conductivity $\sigma_{\text{SiO}_2}=  10^{-8}~\text{S/m}$, while the PBS 1X is highly conductive, with $\epsilon_{\text{PBS 1X}}= 78$ and $\sigma_{\text{PBS 1X}}= 0.15~\text{S/m}$. 

\begin{figure*}[h!]
	\centering
	\includegraphics[width=1.0\textwidth,scale=1]{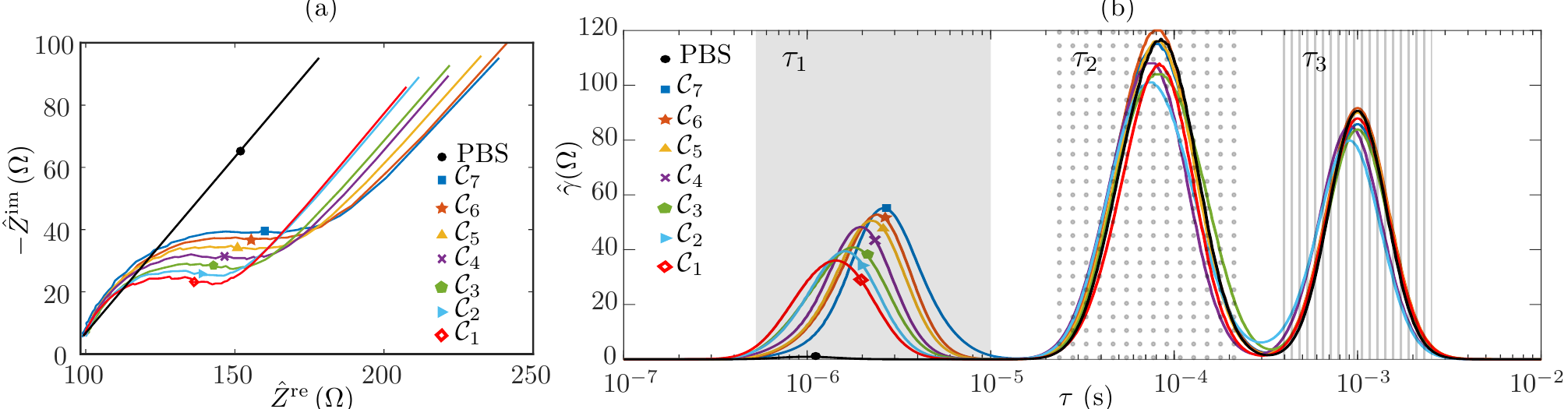}
	\caption{Results of mimic cell suspensions of $\text{SiO}_2$ micro-particles suspended in PBS 1X at different concentrations $\Csp_k$. (a) Nyquist diagram of the electrical impedance measurements. (b) Time-constant-domain spectra represented by the distribution function $\hat{\gamma}(\cdot)$.}
	\label{fig:nyq_sio2}
\end{figure*}

We measured the seven samples $\Csp_w= \{0.10, 0.25, 0.50, 0.75, 1.00, 1.25, 1.50\}$ (wt.\%), for $w=1,2,\dots,7$, and the PBS 1X as reference, using a volume of 50$\mu$L placed on the WE. The real $\hat{Z}^\re$ and imaginary $\hat{Z}^\im$ components of the impedance were measured using $P=8$ periods. As depicted in the Nyquist diagram of Fig.~\ref{fig:nyq_sio2}(a), one can see two main processes involved in EIS measurements. For all the samples, a straight-line-kind trend appears at low frequencies, which is presumably due to the interaction of the WE with the suspensions~\cite{Barbero2008}.
At high frequencies, for the seven suspensions, one can see a semicircle suggesting changes as a function of the concentration due to the interaction among the particles and the PBS 1X, as extensively described in the literature. Also, it is worth to note that all plots tend to 100 $\Omega$ in the real component, linked to the resistance of PBS 1X liquid matrix. However, from the frequency-domain it is difficult to decode precise information about the electrical processes involved in the suspensions.

For each concentration, we transformed the EIS measurements (Fig.~\ref{fig:nyq_sio2}(a)) to its correspondent TCDS given by the distribution function $\hat{\gamma}(\cdot)$ plotted in Fig.~\ref{fig:nyq_sio2}(b). The time-constant spectra distinguish three local maxima for each evaluated sample. In the region of relaxation times $\tau_1 \approx10^{-6}$ s (shadowed area), the Gaussian-like peaks are well differentiated among them, characterizing each concentration by a shift in the value of $\tau_1$. Conversely, for higher time-constant values, $\tau_2 \approx 0.80\times 10^{-4}$ s (dotted area) and $\tau_3\approx 1.00\times10^{-3}$ s (shaded-line area), the maxima are similar in amplitude and location for all samples. 

From the TCDS analysis it is possible to point out three situations about the distribution function. At low relaxation times ($\tau_1$), there is information related to the concentration of micro-particles in suspension~\cite{PRODAN2008}. In the medium relaxation times ($\tau_2)$ region, the consistent presence of local maxima could be associated to changes in the mobility of the surface charges~\cite{SunMorgan2010}. Finally, at higher times ($\tau_3$), the local maxima are due to the effect of the double-layer occurring at the electrode-electrolyte interface~\cite{Chassagne2016}.

\begin{figure}[h!]
	\centering
	\includegraphics[width=0.65\columnwidth,scale=1]{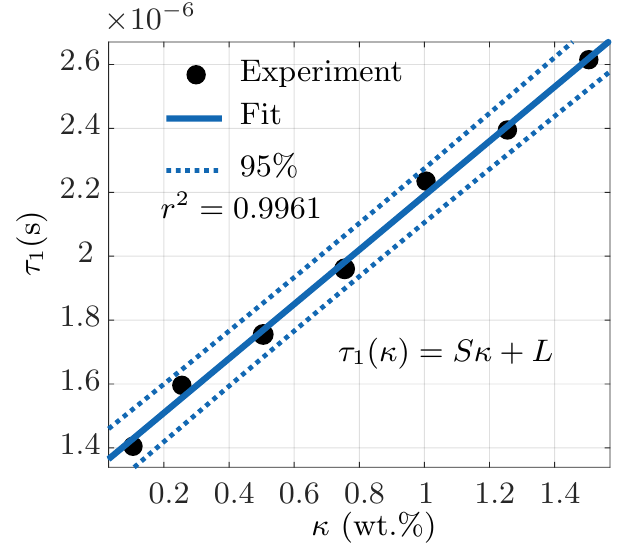}
	\caption{Calibration curve for the relaxation time-constant $\tau_1$ as a function of the concentration $\kappa$ of $\text{SiO}_2$ particles in the suspension. The sensitivity is $S=0.40 \frac{\mu\text{s}}{\text{wt.\%}}$ and $L=1.0\,\mu\text{s}$ is the offset, the goodness of the fit is $r^2$.}
	\label{fig:concentra_1}
\end{figure}

An exhaustive analysis is focused on the low relaxation times region, where one could obtain useful information about the concentration of micro-particles in suspension. The TCDS on Fig.~\ref{fig:nyq_sio2}(b), within the range from $10^{-7}$ to $10^{-5}$ s, shows a shift of the maxima to the right towards higher $\tau_1$ values as the concentration increases. By inspection, the TCDS could serve as a measurement method given the retrieved relaxation times and the concentration of the particles. To this end, in Fig.~\ref{fig:concentra_1} we plot the time-constants $\tau_1$ as a function of the concentration $\kappa$ (wt.\%). The plot shows experimental data (black dots) and the linear model that best fits them (solid line), that in a least-squares sense is $
\tau_1(\kappa)=S\kappa+L$, being $S=0.40 \frac{\mu\text{s}}{\text{wt.\%}}$ the sensitivity and $L=1.0\,\mu\text{s}$ the offset, the goodness of the fit is $r^2=0.9961$. The resolution in $\tau$ is $0.02\times 10^{-7}$ s, given by the frequency resolution of the impedance measurement as well as by the DRT algorithm. The limit of detection for $\kappa$ using the TCDS method is $0.01\,\text{wt.\%}$. 

\subsection{TCDS analysis of HeLa cell suspensions and  impedance parametric model}

Once we calibrated the TCDS method, we applied the TCDS method to asses the electrical properties of biological cell suspensions composed by human cervical cancer cells of the HeLa cell line.
\subsubsection{Cell line and growth conditions}
The HeLa cell line was cultured in Dulbecco's Modified Eagle Medium (DMEM, Gibco, Life Technologies Inc., USA) supplemented with 7\% fetal bovine serum (FBS) (Biowest), 100 U/mL penicillin, 100 $\mu$g/mL streptomycin and 1 mM sodium pyruvate. Cells were maintained in $25~\text{cm}^2$ flask at $37~^{\circ}$C with 5\% CO2 in a humidified atmosphere. 
The culture media was removed and discarded from the flask, and the adherent monolayer of HeLa cells was washed gently with 1 mL of Trypsin-EDTA (0.05-0.05\%) to remove the solution. After 1 mL of Tripsin-EDTA (0.05-0.05\%) was added and incubated for 5-6 min, the detached cells appeared rounded and refractile under microscope. Finally, 3 mL of DMEM complemented was added, the cell suspension was transferred to the tube, and then it was centrifuged at 800 rpm for 5 min. Afterward, the supernatant was removed and the cell pellet was resuspended in pre-warmed DMEM supplemented medium. The number of cells per mL was measured using hemocytometer and trypan blue exclusion. As a result, we prepared $k=1,2,3,4$ concentrations $\Psp_k$, of HeLa cells suspended in PBS 1X, each containing a certain number of cells in a volume of 50 $\mu$L.

\begin{figure*}[h!]
	\centering
	\includegraphics[width=1.0\textwidth,scale=1]{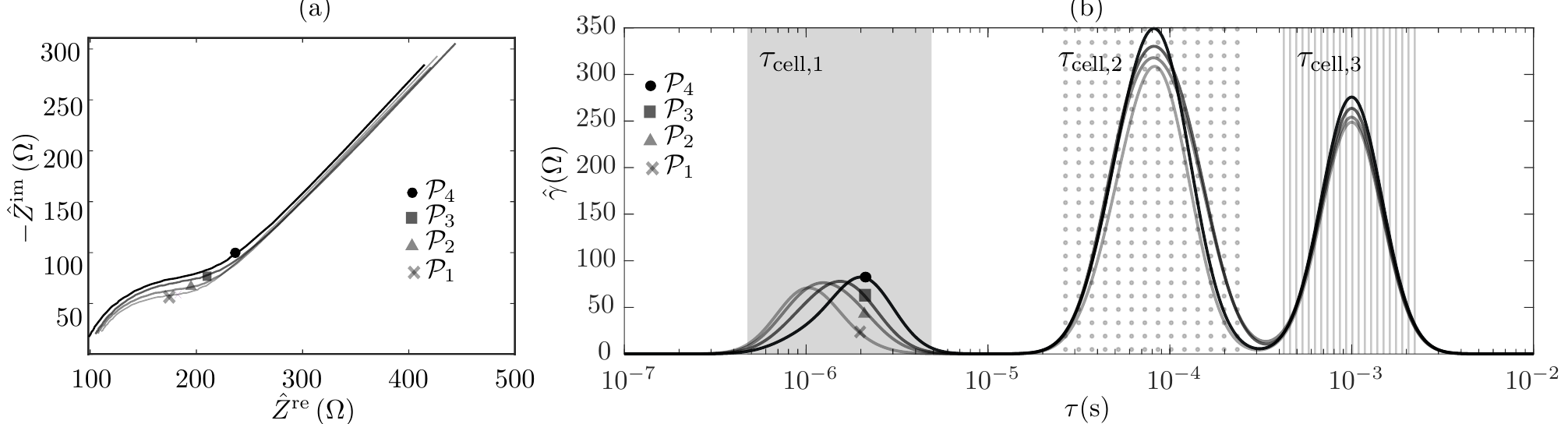}
	\caption{Results of HeLa cells suspended in PBS 1X at different concentrations $\Psp_k$. (a) Nyquist diagram of the electrical impedance measurements. (b) Time-constant-domain spectra represented by the distribution function $\hat{\gamma}(\cdot)$.}
	\label{fig:nyq_cells}
\end{figure*}
\subsubsection{TCDS analysis for sensing HeLa cells}
The impedance measurements were conducted in the frequency range from 1 kHz to 1 MHz using $P=8$ periods of the multisine input/output signals and five experiments to guarantee repeatability. Fig.~\ref{fig:nyq_cells}(a) shows the Nyquist diagrams of the real $\hat{Z}^\re$ and imaginary $\hat{Z}^\im$ components of the measured impedance. As in the case of the suspensions of $\text{SiO}_2$ particles, two processes appear in the frequency-domain response of the cell suspensions. 

At the right side of the Nyquist plot (low frequencies), the straight-line-kind behavior could be related, as already mentioned, to the electrode-electrolyte interface. Meanwhile, at high frequencies, semicircles are described for each cell suspension with a very similar change as that obtained for the mimic suspensions. In  Fig.~\ref{fig:nyq_cells}(b), we show the TCDS analysis in terms of the retrieved $\hat{\gamma}(\cdot)$ function for each cell suspension. As expected, the TCDS clearly decode the underlying processes involved in the electrical response. Interestingly, the maxima values of the Gaussian-like peaks shift in the range of $\tau_{\text{cell},1} \approx10^{-6}$ s (shadowed area). This situation is due to the presence of a large number of cells within the sample, that conduct to a high effective resistance, which in turn is proportional to the relaxation time $\tau_{\text{cell},1:k}$, for $k=1,2,3,4$. Hence, the larger the concentration of cells, the higher the value of the characteristic relaxation time-constant. In the other hand, at higher time-constant values, $\tau_{\text{cell},2}\approx 10^{-4}$ (dotted area) and $\tau_{\text{cell},3}\approx 10^{-3}$ (shaded-line area), the behavior is similar among the four cell suspensions. The presence of maxima therein agrees with the TCDS analysis of the $\text{SiO}_2$ particles (see Fig.~\ref{fig:nyq_sio2}(b)). It is thus possible to elucidate the effects of surface charges and electrode-electrolyte interface, for  $\tau_{\text{cell},2}$ and $\tau_{\text{cell},3}$, respectively. For our purposes in sensing cell suspensions, the high relaxation times $\tau_{\text{cell},2}$ and $\tau_{\text{cell},3}$ can be disregarded. In the third column of Table~\ref{tab2}, we summarize the retrieved constants $\tau_{\text{cell},1}$ for the cell suspensions.

\begin{figure}[t]
	\centering
	\includegraphics[width=1.0\columnwidth,scale=1]{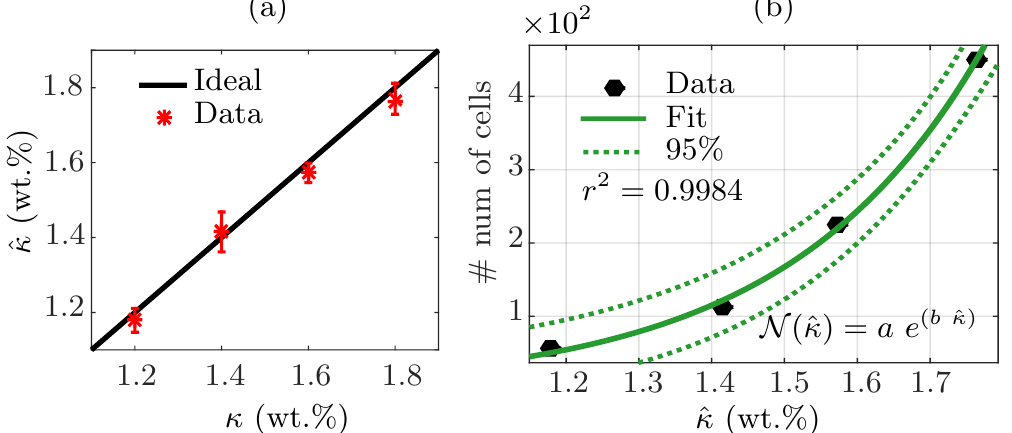}
	\caption{Results of sensing HeLa cell supensions. (a) Scatter plot of the relationship among the estimated $\hat{\kappa}$ and the true $\kappa$ concentrations in (wt.\%). (b) Estimation of the number of cells in 50 $\mu$L as a function of the estimated concentration.}
	\label{fig:dispersion}
\end{figure}

With the calibration curve shown in Fig.~\ref{fig:concentra_1} and its relationship with the TCDS, we estimated the concentration $\hat{\kappa}$ in wt.\% of the HeLa cell suspensions. In Fig.~\ref{fig:dispersion}(a), we show a scatter plot of the estimated and true concentrations. The solid line refers to the ideal relation among such quantities; meanwhile, the asterisk symbols stand for the predicted values and the bars denote the uncertainty of the prediction. From Fig.~\ref{fig:dispersion}(a), it is possible to see that the model well predicts the concentrations for cell suspensions with a correlation coefficient of 0.9939. The fourth column on Table~\ref{tab2} shows the values of $\hat{\kappa}$ and their associated uncertainties. As a metric of performance, we computed the root-mean-square error, $\text{RMSE} = N^{-1}\sqrt{\sum_{k=1}^{N}\left(\hat{\kappa} - \kappa\right)^2}$, with $\hat{\kappa}$ the estimated and $\kappa$ the true concentration, thus leading to $\text{RMSE}=0.0237$ wt.\%. These results allow us to demonstrate that TCDS can be used to detect and quantify cell suspensions with high accuracy.

\begin{table}[h!]
	\caption{HeLa cell samples and their concentration along with the retrieved results using the TCDS:  relaxation times, estimated concentration in weight and number of cells in 50 $\mu$L.}
	\begin{tabular}{cccccc|}
		\hline
		\textbf{Sample} &  $\kappa$~(wt.\%) & $\tau_{\text{cell},1}\times 10 ^{-6}$~(s)  & $\hat{\kappa}$~(wt.\%) & \# of cells\\
		\hline
		$\Psp_1$ & 1.20 & $1.49$ & $1.18 \pm 0.03$& 56 \\
		
		$\Psp_2$ & 1.40 & $1.83$ & $1.41 \pm 0.05$& 112\\
		
		$\Psp_3$ & 1.60 &  $1.95$ & $1.57 \pm 0.02$& 225 \\
		
		$\Psp_4$ & 1.80 &  $2.28$ & $1.76 \pm 0.04$& 450\\
		\hline
	\end{tabular}
	\label{tab2}
\end{table}

Furthermore, as the concentration of the suspensions is related to the number of cells, it is possible to obtain an analytic relationship for these two parameters. In Fig.~\ref{fig:dispersion}(b), we depict the evolution of the number of cells (see Table~\ref{tab2}, fifth column) and the estimated concentration $\hat{\kappa}$. The plot shows the data (symbols), the model that best fits (solid line), and the 95\% confidence intervals in the prediction (dotted lines). From the plot, one can see an exponential-like trend, suggesting the prediction model $\Nsp(\hat{\kappa})=a~ \exp {(b~\hat{\kappa})}$, with the parameters $a=0.604$ and $b=3.749$. The goodness of the fit is $r^2=0.9984$ and the 95\% confidence intervals conducts to a maximum uncertainty of 30 cells.

\subsubsection{Parametric model derivation for cell suspensions}
Let us recall the model structure proposed in~\eqref{eq:model_drt}. According to the results shown in Fig.~\ref{fig:nyq_cells}(b), TCDS decoded $n=3$ $RC$ circuits for each suspension of HeLa cells, thus the impedance model structure can be written as
\begin{align}
H_{\text{TCDS}}(j\omega)= R_{\infty} + \sum_{i=1}^{n}\frac{\Rsp_i}{1+j\omega\tau_{\text{cell},i}} ,
\label{eq:mod_str_drt}
\end{align}
where the time-constants $\tau_{\text{cell},i}$ are straightforward given by the TCDS. For comparison purposes, we contrasted the model in~\eqref{eq:mod_str_drt} with the rule-of-thumb equivalent circuit (EC) considering the capacitance of the electrodes $C_{\text{e}}$ as a CPE. The procedure was done by fitting the raw impedance data to the models~\eqref{eq:mod_str_drt} and~\eqref{eq:zcir} using non-linear-least-squares (NLLS) and the Levenberg-Marquardt algorithm. As a metric of performance, we computed the fitting errors in optimal solution by means of a quadratic parametric error $e^2_p$ between the measurements and the model. As a result, the TCDS-based model exhibits superior accuracy than the EC model, being the maximum error of 0.014 and 0.329, respectively.

By identifying the model structure with TCDS, one can only be interested on the electrical response of the cells. Hence, the last two terms in~\eqref{eq:mod_str_drt} can be neglected, and according to the relation shown in Fig.~\ref{fig:concentra_1}, it follows that
\begin{align}
H_{\text{TCDS,cell}}(j\omega) = R_{\infty} + \frac{\Rsp}{1+j\omega(S\kappa + L)},
\end{align}
where clearly, the impedance is a function of the concentration $\kappa$. With this approach, the model order is thus reduced and its parameters can be easily estimated with a variety of algorithms. This proposed equivalent model intends to serve as a promising tool for further applications for cell counting and monitoring.

\section{Conclusion}\label{sec:conclusions}
In this paper, we introduced an attractive method to characterize and sense biological cell suspensions using impedance measurements and the so-called time-constant-domain spectroscopy (TCDS). The method is based on low-uncertainty impedance measurements and the DRT model. This proposal has two main advantages comparing with state-of-the-art methods for sensing cellular suspensions. First, TCDS can reveal an electrical fingerprint for the suspensions to decode the underlying electrical processes; and second, it can sense biological cells using non-sophisticated instrumentation and avoiding the functionalization of the electrodes. We showed measurements using nominal electrical networks for validation, mimic samples for calibration, and human cell suspensions, for which, the TCDS method was able to sense the concentration of HeLa cells with a maximum error of 0.05 wt.\%. Moreover, based on TCDS, a simple model that relates the impedance-dynamics with the concentration of the suspensions was proposed. To go further than the classical impedance models, TDCS could be a fast and accurate tool for measuring cell suspensions, as well as for assessing and modeling bioelectrical phenomena.
The TCDS could be applied for real-time monitoring of cell
growth, characterizing properties of cell lines, measuring the mobility of cells, and assessing double-layer interfaces.

\section*{Acknowledgment}
This work was partially supported by grants UNAM-PAPIIT IT100518, UNAM-PAPIME PE115319.

\ifCLASSOPTIONcaptionsoff
  \newpage
\fi

\bibliographystyle{IEEEtran}


\begin{IEEEbiographynophoto}{Roberto G. Ram\'irez-Chavarr\'ia} received the BSc degree in Computer Science engineering in 2013, and the MSc and PhD degrees in Electrical Engineering with an Instrumentation profile, from Universidad Nacional Autónoma de México (UNAM) in 2015 and 2019, respectively. Since 2013 he has been a part-time Lecturer at Facultad de Ingeniería, UNAM. In 2019 he held a postdoctoral position at Instituto de Ingeniería, UNAM, and since 2020 he is an Associate Professor in the Electro-Mechanics Department of Instituto de Ingeniería, UNAM. He has been a visitant researcher at Universities in Belgium, Sweden and France in 2017, 2018 and 2019, respectively. His research areas include biosensors and bioelectronics, electrochemistry, and signal processing and artificial intelligence.
\end{IEEEbiographynophoto}

\begin{IEEEbiographynophoto}{Celia Sánchez-Pérez} received the BSc degree in mechanical and electrical engineering from Universidad Nacional Autónoma de México (UNAM), Ciudad de México in 1996. She received the M.S. degree in 1997 in optics, optoelectronic and microwaves and the Ph.D. degree in optoelectronics in 2000 from the Institut National Polytechnique de Grenoble (INPG) France. In 2001 she pursued a postdoctoral research position at the INPG in France. In 2002 she was an associated researcher and since 2010 she has been a full time researcher with the Instituto de Ciencias Aplicadas y Tecnología at Universidad Nacional Autónoma de México and formed part of the National Research System from the National Science and Technology Research Council. Her research interests include biomedical sensors and biophotonics.
\end{IEEEbiographynophoto}

\begin{IEEEbiographynophoto}{Luisa Romero-Ornelas} recieved the BSc degree in Biology from Escuela Nacional de Ciencias Biológicas at Instituto Politécnico Nacional-IPN (2019). She is currently pursuing a MSc degree in Biomedical Sciences and Molecular Biotechnology at the ENCB of IPN. She has participated as a fellow in the construction, development and dissemination of a science-communication workshop, where microfluidic devices were manufactured for detecting heavy metals in contaminated water.		
\end{IEEEbiographynophoto}

\begin{IEEEbiographynophoto}{Eva Ramón-Gallegos} majored in Pharmaceutical Biology Chemistry from Veracruz University (1994), MSc in Cytophatology from Instituto Politécnico Nacional-IPN (1997), and PhD in Biomedicine from the Centro de Estudios Avanzados of IPN (CINVESTAV) (2000). She is a Researcher and Professor of Instituto Politécnico Nacional in Mexico since 2001 and Environment Cytopatology laboratory responsible in the Morphology Department of the Escuela Nacional de Ciencias Biologicas of IPN. Her main investigation areas are: Photodynamic Therapy, Nanotechnology and Cancer. She is member of the National Research System of Mexico (Sistema Nacional de Investigadores) level 2 from 2001 to date and is member of International Photodynamic Association from 2012. She has 2 patent and 4 patent applications (1 international and 3 national).
\end{IEEEbiographynophoto}

\end{document}